\documentstyle[preprint,aps,epsf]{revtex}
\tightenlines

\def\simlt{\rlap{\lower 3.5 pt\hbox{$\mathchar \sim$}}\raise 1pt \hbox {$<$}}
\def\simgt{\rlap{\lower 3.5 pt\hbox{$\mathchar \sim$}}\raise 1pt \hbox {$>$}}

\begin{document}
\draft

\title{Kaon $B$ parameter from quenched Lattice QCD}

\author{JLQCD Collaboration\\
$^1$S.~Aoki, $^2$M.~Fukugita, $^3$ S.~Hashimoto,
$^{1,4}$N.~Ishizuka, $^{1,4}$Y.~Iwasaki, $^{1,4}$K.~Kanaya, $^5$Y.~Kuramashi,
$^5$M.~Okawa, $^1$A.~Ukawa, $^{1,4}$T.~Yoshi\'{e}}

\address{
$^1$Institute of Physics, University of Tsukuba, 
Tsukuba, Ibaraki-305, Japan \\
$^2$Institute for Cosmic Ray Research, University of Tokyo,
Tanashi, Tokyo 188, Japan \\
$^3$Computing Research Center, High Energy Accelerator Research 
Organization(KEK), Tsukuba, Ibaraki 305, Japan \\
$^4$Center for Computational Physics,  University of Tsukuba,
Tsukuba, Ibaraki 305, Japan \\
$^5$Institute of Particle and Nuclear Studies, High Energy Accelerator 
Research Organization(KEK), Tsukuba, Ibaraki 305, Japan \\
}

\date{\today}

\maketitle

\begin{abstract}
We present results of a large-scale simulation for the Kaon $B$ 
parameter $B_K$ in quenched lattice QCD with the Kogut-Susskind quark action.
Calculating $B_K$ at 1\% statistical accuracy for seven values of lattice 
spacing in the range $a\approx 0.24-0.04$~fm on lattices up to 
$56^3\times 96$, we verify a quadratic $a$ dependence of $B_K$ 
theoretically predicted.  Strong indications are found that, with our 
level of accuracy, $\alpha_{\overline{MS}}(1/a)^2$ terms arising from 
our one-loop matching procedure have to be included in the continuum 
extrapolation. 
We present $B_K$(NDR, 2 GeV)=0.628(42) as our final value, as obtained
by a fit including the $\alpha_{\overline{MS}}(1/a)^2$ term.
\end{abstract} 


\narrowtext

The knowledge of the kaon $B$ parameter $B_K$ 
\begin{eqnarray}
B_K &=& \frac{
\langle \bar K^0 | \bar s\gamma_\mu (1-\gamma_5)d\cdot
\bar s\gamma_\mu (1-\gamma_5)d | K^0\rangle
}
{
\frac{8}{3}
\langle \bar K^0 | \bar s\gamma_\mu\gamma_5 | 0\rangle
\langle \bar 0 | \bar s\gamma_\mu\gamma_5 | K^0\rangle
} ,
\end{eqnarray}
is imperative to extract the CP violation parameter of 
the Cabibbo-Kobayashi-Maskawa matrix from experiment.
Work has been continued for a decade to determine
this parameter with lattice QCD\cite{OldWork} using both  
Wilson and Kogut-Susskind quark actions.  Calculations with the 
latter have the advantage\cite{STAG} that the correct chiral 
behavior of the matrix element is ensured by $U(1)$ chiral symmetry. 
Nonetheless, previous studies with this 
action\cite{STAG,IFMOSU,kilcupfull,KGS} 
have not yielded a definitive result for the matrix element.

A major difficulty, uncovered in Ref.~\cite{STAG}, 
is the presence of a large scaling violation in $B_K$, which 
renders a reliable extrapolation to the continuum limit nontrivial.
Whereas the scaling violation is theoretically expected to be 
$O(a^2)$\cite{sharpe1} with the Kogut-Susskind 
action,
simulations so far\cite{STAG,IFMOSU,KGS} could not confirm
it due to large statistical errors.

Another problem concerns systematic uncertainties in 
the renormalization factors needed to match the lattice result
to that in the continuum. While an earlier study\cite{IFMOSU} 
found that one-loop perturbation theory is reasonably accurate, 
the problem of the systematic error associated with 
renormalization has not been fully explored yet.

In order to resolve these problems, we have carried out a 
large-scale simulation for $B_K$ with the Kogut-Susskind quark action 
in quenched lattice QCD. 
In this article we report on the continuum limit 
of $B_K$, expounding
the crucial points of our simulations and analysis.

The parameters employed in our simulations 
are summarized in Table~\ref{tab:run}.
In order to study the continuum limit, seven values of 
the inverse gauge coupling constant $\beta=6/g^2$ spanning the range 
$\beta=5.7-6.65$ are chosen for the simulations, corresponding to the 
lattice spacing $a\approx 0.24-0.04$~fm.  We set the physical 
scale of lattice spacing by $\rho$ meson mass in the $VT$ channel.  
The physical lattice size is kept approximately constant 
at $La\approx 2.3-2.5$~fm in order to distinguish scaling violation 
effects from those of finite lattice.
Finite-size effects are examined separately 
at $\beta=6.0$ and 6.4, varying the lattice size 
over the range $La\approx 1.8-3.1$ fm.
Numerical simulations have been carried out
on the Fujitsu VPP500/80 supercomputer at KEK.

We employ both gauge-invariant and noninvariant four-quark 
operators\cite{IFMOSU}, which differ by an insertion 
of gauge link factors connecting the quark fields 
spread over a $2^4$ hypercube.  The bare lattice operators are mean-field 
improved through a replacement
$\chi\rightarrow\sqrt{u_0}\chi$ for the quark field and  
$U_\mu \rightarrow u_0^{-1} U_\mu$ for the gluon field, 
where $u_0 =P^{1/4}$\cite{LM}
$P$ being the average value of the plaquette.

The matching of $B_K$ between lattice and continuum is made  
in the following way.  We first correct lattice values of $B_K$ 
by the one-loop renormalization factor\cite{IS,SP} 
evaluated with the $\overline{MS}$ coupling $\alpha_{\overline{MS}}(q^*)$ 
at a matching scale $q^*=1/a$\cite{Ji,GBS} to obtain the continuum operator
$B_K({\rm NDR},q^*)$ 
renormalized in the $\overline{MS}$ 
scheme with the naive dimensional regularization(NDR).
The continuum value at a physical scale $\mu =$ 2 GeV
is then obtained via a 2-loop running of the continuum
renormalization group starting from $B_K({\rm NDR}, q^*)$:
\begin{eqnarray}
B_K({\rm NDR}, \mu ) & = &
\left[ 
1- \frac{\alpha_{\overline{MS}}(\mu)}{4\pi}
\frac{\gamma_1\beta_0-\gamma_0\beta_1}{2\beta_0^2} 
\right]^{-1}
\left[ 
1- \frac{\alpha_{\overline{MS}}(q^*)}{4\pi}
\frac{\gamma_1\beta_0-\gamma_0\beta_1}{2\beta_0^2}
\right] \nonumber \\
&\times &
\left[\frac{\alpha_{\overline{MS}}(q^*)}{\alpha_{\overline{MS}}(\mu)}
\right]^{-\gamma_0/2\beta_0}
B_K({\rm NDR},q^*)
\end{eqnarray}
where $\beta_0=11$, $\beta_1 =102$, $\gamma_0 =4$ and $\gamma_1 = -7$
\cite{BJW} are the $N_f=0$ 
quenched values for the renormalization group coefficients.
This procedure leaves an uncertainty of 
$O(\alpha_{\overline{MS}}(q^*)^2)$ in $B({\rm NDR},q^*)$ arising 
from the use of one-loop renormalization factors\cite{KGS,Onogi}. 
 
The coupling constant $\alpha_{\overline{MS}}(q^*)$ needed in the matching 
factor is obtained once the $\Lambda_{\overline{MS}}$ is specified.
To estimate this, we start from 
$\alpha_P$\cite{alphap} defined by  
$ - \log P =4\pi/3\alpha_P(3.40/a)(1-1.19\alpha_P)$, 
and calculate $\Lambda_{\overline{MS}}=0.625 \Lambda_P$ from 
$\alpha_P(3.40/a)$, where the 3-loop correction term
is included. 
The value of $\Lambda_{\overline{MS}}$ estimated in this way, 
however, suffers from 
scaling violation. 
We therefore extrapolate the results at our seven values of $\beta$ 
quadratically in $m_\rho a$ to the continuum limit, 
finding $\Lambda_{\overline{MS}}=232(4)$~MeV.
We then take $\Lambda_{\overline{MS}}=230$~MeV,  
and calculate the $\overline{MS}$ running coupling to 3-loop accuracy,
which is used throughout our analyses to minimize additional scaling 
violation entering into the $B_K$ calculation. 

In our simulations  
gauge configurations are generated with the 5-hit 
heatbath algorithm, and $B_K$ is calculated 
at every 1000($\beta = 5.7$), 2000($\beta\leq 6.0$) 
or 5000($\beta\geq 6.2$) sweep intervals.
Our main results are based on calculations at four values of 
degenerate strange and down quark mass $m_qa$, equally spaced in the 
interval given in Table~\ref{tab:run}.

Lattice values of $B_K$ are calculated from the 3-point 
Green function of the four-quark operator at time $t$ 
with two kaons created at the temporal edges of the lattice, 
divided by the vacuum saturation of the same operator.
Eight wall sources corresponding to the corners of a spatial cube 
are employed to construct a quark-antiquark propagator combination 
such that only the pseudoscalar meson in the Nambu-Goldstone channel 
propagates\cite{8wall}.
Quark propagators are calculated with the Dirichlet boundary condition 
in time and the periodic boundary condition in space.
Gauge configurations are fixed to the Landau gauge.

The fitting interval to extract $B_K$ from
the Green function is chosen so that 
the minimum time $t_{min} a $ is approximately constant at 
$t_{min}a\approx  1.4-1.5$ fm for all values of $\beta$.
The resulting $B_K$ changes less than $\pm$0.3 \% for all $m_qa$ and
$\beta$ under a variation of $t_{min}$ by $\pm 2$.

At each value of $\beta$ lattice results are interpolated in $m_q a$ 
with the formula suggested by chiral perturbation theory\cite{sharpe2},
\begin{equation}
B_K = B [ 1- 3 c\cdot m_q a \log (m_q a) + b\cdot m_q a ] .
\end{equation}
The physical value of $B_K$ is obtained at half the strange 
quark mass $m_s a/2$, estimated from experiment,
$m_K/m_\rho = 0.498/0.770$.

We present our results in Table~\ref{tab:bk}.
The errors  
are estimated by a single elimination jackknife procedure.
Our statistical error is small, being 0.1\% at $\beta=5.7$ and 
gradually increasing to 1.2\% at $\beta=6.65$.  
At $\beta$ = 6.0 and 6.4, three spatial sizes are examined for 
a finite-size study.  Some size dependence of 
order 2\% is seen below the spatial size $La\approx $2.0~fm 
at $\beta=6.4$, but the magnitude decreases to less than 0.5\% for 
$La\simgt 2.2$ fm at both values of $\beta$.  We have made our
main runs with a spatial size larger than $La\approx 2.3$~fm,
thus expecting  
finite-lattice corrections being smaller than the statistical error.

\begin{figure}[bt]
\centerline{\epsfxsize=8.5cm \epsfbox{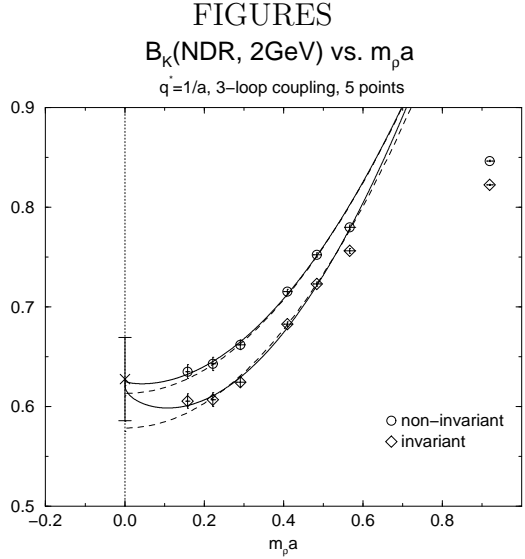}}
\caption{Gauge non-invariant (circles) and invariant (diamonds)
$B_K$({\rm NDR}, 2GeV) as a function of
$m_\rho a$, together with a simultaneous fit for the two operators
including $\alpha^2$ term (solid lines)
and separate fits quadratic in $a$(dashed lines)
to the five pairs of data points for $\beta\geq 5.93$.}
\label{fig:bkmro}
\end{figure}

We present $B_K ({\rm NDR}, 2 {\rm GeV})$ 
as a function of $m_\rho a $ in Fig.~\ref{fig:bkmro} both for
gauge noninvariant (circles) and invariant (diamonds) operators.
The five points below $m_\rho a\approx 0.6 (\beta\geq 5.93)$ 
are consistent with the $O(a^2)$ scaling behavior 
theoretically expected\cite{sharpe1}. 
Toward large lattice spacings, however, we observe a change of 
curvature from a positive to a negative sign. 
At an intermediate range $m_\rho a\approx 0.6-0.3$ ($\beta=5.85-6.2$) 
a cancellation among the $a^2$ and higher order terms 
conspire to yield an apparently linear dependence of $B_K$.  
This is the linear behavior 
we observed at an early stage of our work\cite{JLQCD95}.  The later
result at a smaller lattice spacing $m_\rho a \approx 0.22 (\beta=6.4)$ 
gave a first indication of  
an $O(a^2)$ behavior\cite{JLQCD96}; 
this is now confirmed by the calculation at a yet smaller 
lattice spacing 
$m_\rho a\approx 0.16 (\beta=6.65)$ given in this paper.

In our preliminary report\cite{JLQCD96} we took a naive approach
to estimate the  continuum $B_K$, simply by applying 
a polynomial fit assuming $O(a^2)$ dependence.
A fit of the five points above $\beta=5.93$
with the form $B_K=c_0+c_1(m_\rho a)^2$,
shown by the dashed lines in Fig.~\ref{fig:bkmro},
gives a value at the continuum
$B_K$(NDR, 2 GeV)=0.616(5) for the gauge
noninvariant operator, and 0.580(5) for the invariant one,
the average of the two being 0.598(5).

\begin{figure}[bt]
\centerline{\epsfxsize=8.5cm \epsfbox{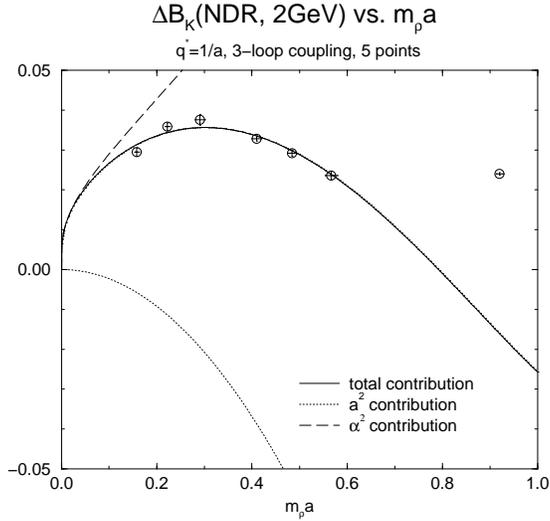}}
\caption{Difference of $B_K$({\rm NDR}, 2GeV) between
gauge noninvariant and invariant operators
as a function of $m_\rho a$.
The solid line represents a fit with $a^2$ and $\alpha^2$ terms,
while the dotted (dashed) line 
is the contribution from the $a^2$ ($\alpha^2$) term.}
\label{fig:difBK}
\end{figure}

An obvious problem with this analysis is that the two operators yield 
different values.
We recall that $B_K$ for the two operators, and hence also 
their difference, should receive not only $O(a^2)$ scaling violation 
but also $\alpha_{\overline{MS}}(q^*)^2$ errors from the matching 
procedure.
Figure~\ref{fig:difBK} plots the difference as a function of  $m_\rho a$ 
(numerical values given in Table~\ref{tab:bk}).
Small errors of 3--4\% result from a strong correlation between 
the matrix elements of the two operators.  
We find that the difference can be fitted by the form  
$b_1 (m_\rho a)^2 + b_2\alpha_{\overline{MS}}(q^*)^2$: 
employing five data points for $m_\rho a\simlt 0.5$ we  
obtain $b_1=-0.23(2)$ and $b_2=1.73(5)$ for $\chi^2/{\rm d.o.f}=2.2$. 
The solid line indicates the fit, and the 
others show the breakdown into the $a^2$(dotted line) and 
$\alpha^2$(dashed line) 
contributions.
A fit allowing a constant $b_0$ yields a value of 
$b_0$ vanishing within $2\sigma$:
$b_0=-0.032(16)$, $b_1=-0.44(11)$ and $b_2=3.4(8)$.
These results strongly indicate that the decrease of the difference 
toward small lattice spacings seen in Fig.~\ref{fig:difBK} is 
actually an $\alpha_{\overline{MS}}(q^*)^2$ effect.

Encouraged by this analysis 
we attempt to fit the five points at $\beta\ge 5.93$
simultaneously for both operators including their correlations 
with the form
$B_K^{\rm non-inv}=c_0+c_1a^2+c_2\alpha_{\overline{MS}}(q^*)^2$
and 
$B_K^{\rm inv}=d_0+d_1 a^2+d_2\alpha_{\overline{MS}}(q^*)^2$.
This yields $c_0=0.67(6)$ and $d_0=0.71(7)$, and hence we
impose the constraint  $c_0=d_0$ in our final fit.
In the continuum limit the fit (solid lines in Fig.~\ref{fig:bkmro}) gives
$B_K$ (NDR, 2 GeV) = 0.628(42) with $\chi^2$/d.o.f=1.37. 
The error is roughly ten times the one from the naive quadratic fit. 
This large error reflects uncertainties of the coefficient of 
the $\alpha^2$ terms:
$c_2=-0.5(2.0)$ and $d_2=-2.2(2.0)$. 
The difference, however, is well constrained; $c_2-d_2=1.7$ 
agrees well with $b_2=1.73$ obtained above.

We find larger coefficients $c_2=-1.0(4.2)$ and 
$d_2=-4.3(4.2)$ when $q^*=\pi/a$ is used,
or $c_2=1.6(1.5)$ and $d_2=-3.2(1.5)$
if mean-field improvement is not made for the operators. 
This supports the tadpole argument of Ref.~\cite{LM}.

The final value depends only weakly on our choice of 
$\Lambda_{\overline{MS}}=230$~MeV: {\it e.g.,}
$B_K$(NDR, 2~GeV)=0.627(42) 
for  $\Lambda_{\overline{MS}}$=220~MeV and 0.628(41) for 240~MeV.

As our final value of $B_K$ in the continuum limit, 
we adopt the fit including the $\alpha^2$ term,
\begin{equation}
B_K ({\rm NDR, 2 GeV}) = 0.628 \pm 0.042,
\label{eq:final}
\end{equation}
which includes a systematic error from the 2-loop uncertainty.
The size of the quoted error is 6.6\%, which roughly equals 
$3\times\alpha_{\overline{MS}}(q^*=1/a)^2$
at our smallest lattice spacing $1/a=4.87{\rm GeV}$ at $\beta =6.65$ 
where $\alpha_{\overline{MS}}(4.87{\rm GeV})=0.147$.
This magnitude of error is unavoidable, 
even with 1\% statistical accuracy at each $\beta$ achieved in 
our simulation, 
unless a two-loop calculation is carried out
for the lattice renormalization.

Our final result is 
consistent with the JLQCD value obtained using the 
Wilson quark action,
$
B_K({\rm NDR, 2 GeV})=0.562\pm 0.064 ,
$
in which the operator mixing problem is solved
non-perturbatively with the aid of chiral Ward identities\cite{JLQCD97}.

One of the systematic errors not taken into account in our final
result (\ref{eq:final}) is the effect of non-degenerate
strange and down quark masses $m_s\ne m_d$.
Analyzing this problem is difficult within quenched QCD since the 
chiral limit $m_d\to 0$ with $m_s\ne 0$ is expected to diverge due to 
a quenched chiral logarithm\cite{QChPT}.
Our attempt at a verification of the logarithmic divergence is also 
inconclusive: our results for non-degenerate quarks can be fitted 
quite well either with or without the singular term.  
At this stage we are not able to quote the
magnitude of error due to the use of degenerate quark masses.

Finally our quoted error does not include effects of  
sea quarks.  Preliminary attempts  
suggest that the quenching error may not exceed 5\% or 
so\cite{IFMOSU,kilcupfull,full}. 
More extensive efforts, however, are clearly needed to estimate
dynamical quark contributions to the  $B_K$ parameter.  
Full QCD simulations should also enable us to answer the issues with 
non-degenerate quark masses.  Carrying out such calculations 
represent the final step toward 
the first-principle determination of the kaon $B$ parameter.

This work is supported by the Supercomputer
Project (No.~97-15) of High Energy Accelerator Research Organization (KEK),
and also in part by the Grants-in-Aid of
the Ministry of Education (Nos~08640349, 08640350, 08640404,
09246206, 09304029, 09740226).



\mediumtext
\begin{table}
\caption{Run parameters}
\label{tab:run}
\begin{center}
\begin{tabular}{llllllllll}
$\beta$ & $m_qa$ & $L^3 T$ & \# conf. & $m_\rho$ & $a^{-1}$(GeV) & $La$(fm) 
& $t_{min}-t_{max}$ & $m_s a/2$ & $P$ \\
\hline
5.7  & 0.02-0.08 & $12^3 24$ & 150 &0.9120(7)& 0.837(6)& 2.83 & 6-16  & 
0.0519(8) & 0.54900 \\
5.85 & 0.01-0.04 & $16^3 32$ & 60  &0.567(13)& 1.36(3) & 2.32 & 10-20 
& 0.0201(9) & 0.57506\\
5.93 & 0.01-0.04 & $20^3 40$ & 50  &0.484(10)& 1.59(3) & 2.48 & 12-26 
& 0.0160(6) & 0.58564\\
6.0  & 0.01-0.04 & $24^3 64$ & 50  &0.410(10)& 1.88(4) & 2.52 & 15-47 
& 0.0125(5) & 0.59374 \\
     &           & $18^3 64$ & 50  &0.413(12)& 1.87(6) & 1.90 & 15-47 
& 0.0127(7) &\\
     &           & $32^3 48$ & 40  &0.383(3) & 2.01(2) & 3.14 & 15-31 
& 0.0109(2) & \\
6.2  & 0.005-0.02 & $32^3 64$ & 40  &0.291(10)& 2.65(9) & 2.39 & 20-42 
&0.00884(57) & 0.61365 \\
6.4  & 0.005-0.02 & $40^3 96$ & 40  &0.222(5) & 3.47(7) & 2.28 & 25-69 
&0.00692(29) & 0.63065 \\
&     & $32^3 96$ & 40  &0.216(7) & 3.57(11)& 1.77 & 25-69 &0.00659(41) & \\
&     & $48^3 96$ & 20  &0.219(4) & 3.52(7) & 2.69 & 25-69 &0.00681(26) & \\
6.65 & 0.004-0.016 & $56^3 96$ & 40  &0.158(4) & 4.87(12)& 2.27 & 36-58 
&0.00511(25) & 0.64912 \\
\end{tabular}
\end{center}
\end{table}

\narrowtext
\begin{table}
\caption{Results for $B_K({\rm NDR, 2GeV})$ at each $\beta$ calculated 
with the matching scale $q^* =1/a$.}
\label{tab:bk}
\begin{center}
\begin{tabular}{ll|ll|l}
$\beta$ & $L^2 T$ &\multicolumn{2}{c|}{ $B_K$(NDR, 2GeV)}&$\Delta B_K$ \\
& & non-invariant & invariant & \\
\hline
5.7  & $12^3 24$ & 0.8464(7)  & 0.8224(7) &0.0240(3)  \\
5.85 & $16^3 32$ & 0.7798(25) & 0.7562(25)&0.0236(11) \\
5.93 & $20^3 40$ & 0.7522(23) & 0.7229(22)&0.0292(8) \\
6.0  & $24^3 64$ & 0.7154(23) & 0.6826(24)&0.0328(5) \\
     & $18^3 64$ & 0.7174(68) & 0.6787(68)&0.0388(12) \\
     & $32^3 48$ & 0.7128(14) & 0.6790(16)&0.0339(8) \\
6.2  & $32^3 64$ & 0.6619(48) & 0.6243(45)&0.0376(14) \\
6.4  & $40^3 96$ & 0.6428(67) & 0.6069(69)&0.0359(10) \\
     & $32^3 96$ & 0.6577(122)& 0.6126(112)&0.0451(24)\\
     & $48^3 96$ & 0.6415(48) & 0.6072(51) &0.0343(11)\\
6.65 & $56^3 96$ & 0.6350(70) & 0.6055(72)&0.0295(10) \\
\end{tabular}
\end{center}
\end{table}

\end{document}